\documentclass[conference]{IEEEtran}
 
\IEEEoverridecommandlockouts
\usepackage{cite}
\usepackage{amsmath,amssymb,amsfonts}
\usepackage{algorithmic}
\usepackage{graphicx}
\usepackage{textcomp}
\usepackage{xcolor}
\usepackage{float}
\usepackage{comment}
\usepackage{lipsum}
\usepackage{caption}
\usepackage{subcaption}
\usepackage{array,multirow,graphicx}
\begin{document}

\title{%CARLA Simulator Camouflaged Adversarial Attacks{\color{red}--how about ``
Mitigation of Camouflaged Adversarial Attacks in Autonomous Vehicles--A Case Study Using CARLA Simulator\\
\vspace{-7mm}
}

\author{Yago Romano Martinez, Brady Carter, Abhijeet Solanki, Wesam Al Amiri, Syed Rafay Hasan, and Terry N. Guo 
\vspace{-6mm}
\thanks{Y.R. Martinez, C. Brady, A. Solanki, W. Al Amiri, S.R. Hasan, and T. Guo are with the College of Engineering, Tennessee Technological University, Cookeville, TN (email: \{yromanoma42,  clbrady43, asolanki42, walamiri, shasan, nguo\}@tntech.edu). }%Y.R. Martinez is with  the Department of Computer Science, Cookeville, TN (email:yromanoma42@tntech.edu). B. Carter, A. Solanki, W. Al Amiri, and S.R. Hasan are with the Department of Electrical and Computer Engineering, Cookeville, TN (email:\{asolanki42, clbrady, walamiri, shasan\}@tntech.edu). T. Guo is with the Center for Manufacturing Research, Tennessee Technological University, Cookeville, TN (email: nguo@tntech.edu). }
%\vspace{-10mm}
}
\begin{comment}

\author{\IEEEauthorblockN{1\textsuperscript{st} Syed Rafay Hasan}
\IEEEauthorblockA{\textit{Department of Computer Engineering} \\
\textit{Tennessee Technological University}\\
Cookeville, Tennessee \\
shasan@tntech.edu}
\and
\IEEEauthorblockN{2\textsuperscript{nd} Abhijeet Solanki}
\IEEEauthorblockA{\textit{Department of Computer Engineering} \\
\textit{Tennessee Technological University}\\
Cookeville, Tennessee \\
asolanki42@tntech.edu}
\and
\IEEEauthorblockN{3\textsuperscript{rd} Brady Carter}
\IEEEauthorblockA{\textit{Department of Computer Engineering} \\
\textit{Tennessee Technological University}\\
Cookeville, Tennessee \\
carterlb422@gmail.com}
\and
\IEEEauthorblockN{4\textsuperscript{th} Thomas Dean Robertson II}
\IEEEauthorblockA{\textit{Department of Computer Science} \\
\textit{Tennessee Technological University}\\
Cookeville, Tennessee \\
tdrobertso42@tntech.edu}
\and
\IEEEauthorblockN{5\textsuperscript{th} Yago Romano Martinez}
\IEEEauthorblockA{\textit{Department of Computer Science} \\
\textit{Tennessee Technological University}\\
Cookeville, Tennessee \\
yromanoma42@tntech.edu}
\and
\IEEEauthorblockN{6\textsuperscript{th} Given Name Surname}
\IEEEauthorblockA{\textit{dept. name of organization (of Aff.)} \\
\textit{name of organization (of Aff.)}\\
City, Country \\
email address or ORCID}
}
\end{comment}

\maketitle
\vspace{-15mm}
\begin{abstract}
%\vspace{-2mm}
 Autonomous vehicles  (AVs) rely heavily on cameras and artificial intelligence (AI) to make safe and accurate driving decisions. However, since AI is the core enabling technology, this raises serious cyber threats that hinder the large-scale adoption of AVs. Therefore, it becomes crucial to analyze the resilience of AV security systems against sophisticated attacks that manipulate camera inputs, deceiving AI models. In this paper, we develop camera-camouflaged adversarial attacks targeting traffic sign recognition (TSR) in AVs. Specifically, if the attack is initiated by modifying the texture of a stop sign to fool the AV’s object detection system, thereby affecting the AV actuators. The attack's effectiveness is tested using the CARLA AV simulator and the results show that such an attack can delay the auto-braking response to the stop sign, resulting in potential safety issues. We conduct extensive experiments under various conditions, confirming that our new attack is effective and robust. Additionally, we address the attack by presenting mitigation strategies. The proposed attack and defense methods are applicable to other end-to-end trained autonomous cyber-physical systems.
 %%Abhijeet, Yago, Carter, Wesam, Rafay, Terry
\end{abstract}

\begin{IEEEkeywords}
Autonomous vehicles (AVs), camera perception, camouflaged adversarial attacks, traffic sign recognition (TSR), and CARLA simulator. 
\end{IEEEkeywords}

\vspace{-4mm}
\section{Introduction}
\vspace{-1mm}
%The transportation landscape is transforming due to the rapid development of autonomous vehicles (AVs). These vehicles promise to enhance road safety, improve mobility access, and reduce transportation costs. 

The transportation landscape is undergoing a significant transformation with the rapid development of AVs, which are becoming an integral part of connected intelligent transportation systems \cite{kloukiniotis2022countering,mahima2024toward,parekh2022review}. AVs hold the potential to enhance road safety, improve mobility access, and reduce transportation costs, revolutionizing the way we travel. This transformation is largely driven by the advanced perception system, which enables AVs to comprehend and navigate the dynamic driving environment.  A key element of AV development is traffic sign recognition (TSR), which is a crucial component that ensures the safe and effective operation of these AVs by identifying and interpreting road signs in real time \cite{sarwatt2024adapting}. TSR involves several stages: the acquisition of traffic sign images, preprocessing, detection, and ultimately the classification of traffic signs \cite{triki2023real}. This process allows AVs to accurately perceive and respond to traffic signals, contributing to safer navigation and adherence to road regulations.
\vspace{-0.5mm}
Unlike other perception tasks that can rely on a fusion of sensor data, such as obstacle detection or lane tracking, TSR predominantly hinges on data from camera sensors \cite{pavlitska2023adversarial}. The AV relies on a camera to capture images of traffic signs, which are then transmitted to an AI model, specifically Deep neural networks (DNNs), for interpretation and decision-making. DNNs have shown remarkable effectiveness in recognizing traffic signs under standard conditions, achieving accuracy rates that exceed human capabilities \cite{soylu2024performance}.

However, the use of DNN in TSR has exposed a vulnerability to adversarial attacks targeting AV sensors. These attacks alter  traffic sign images to generate ``adversarial examples'', which deceive the DNNs, potentially leading to incorrect interpretation of traffic signs and dangerous decisions by AVs. Therefore, ensuring the robustness of AV systems against adversarial sensor attacks, particularly those targeting camera perception, is critical for their secure operation. 

Recent research studies \cite{jia2022fooling,zhong2022shadows} have demonstrated that adversarial attacks, such as modifying signs using stickers or light projections, can cause autonomy stacks to misclassify traffic signs. For instance, these attacks  might cause a stop sign being mistakenly identified as %a yield or 
speed limit sign, potentially causing dangerous situations or accidents. However, these studies have some limitations. First, human drivers can often detect these attacks. Drivers may notice stickers or lighting inconsistencies that do not match the surrounding environment on a sign, limiting the effectiveness of these attacks in semi-autonomous AVs (e.g., Tesla's), where the driver can intervene and regain control. Additionally, these studies do not address scenarios where attacks are undetectable by humans, leaving uncertainty about the risks posed to fully autonomous systems \cite{sato2023wip}. %} \textcolor{red}{Need to check if there are other attacks and mention their limitations .........}. 
 Therefore, there is an urgent need to analyze more sophisticated cyberattacks, those that do not require physical access and can deceive systems by altering objects remotely, to gain critical insights for ensuring AVs safety in the face of evolving cyber threats. Vision  attacks, which manipulate camera inputs, can significantly disrupt object detection systems, leading to potential navigation errors and safety risks. The existing studies have explored the affect of adversarial attacks on the machine learning model capability. However, to the best of authors' knowledge, joint investigation of attack and defense of such attacks in a controlled and realistic AV traffic simulator is not well studied. %While previous studies have explored various aspects of these threats, there is a need for a
%comprehensive analysis focused on camera perception within a controlled simulation environment.

%Recent research has demonstrated attacks, such as adding stickers or dark patches to signs, that cause CAV sign misinterpretation, resulting in potential safety issues. 

%their increasing prevalence on public roads introduces a heightened risk of sophisticated cybersecurity threats. Ensuring the robustness of AV systems against adversarial sensor attacks, particularly those targeting camera perception, is critical for their secure operation.
\vspace{-0.5mm}
In this paper, our research focuses on simulating and analyzing the impacts of adversarial attacks on camera perception, which directly affects object detection and, consequently, the navigation tasks of AVs. Specifically, we aim to:
\vspace{-1mm}
\begin{enumerate}
    \item Evaluate the resilience of AV security systems against camouflage adversarial attacks that manipulate camera inputs.
    \item Assess the effectiveness of defensive measures in mitigating these threats.
\end{enumerate}
\vspace{-1mm}
Our primary research questions are:
\begin{enumerate}
  \vspace{-1mm}
    \item How do adversarial attacks on camera perception affect the object detection capabilities of AVs?
    %\vspace{-2mm}
    \item What detection and/or mitigation approach can be beneficial in such an environment? 
\end{enumerate}

\vspace{-2.8mm}
\section{Simulation Environment}
\vspace{-1mm}

Our research utilizes the CARLA AV simulator \cite{carla}, %a realistic and customizable platform for simulating sensor attacks and defenses. 
which offers a highly customized and controlled environment for conducting experiments and collecting data that would be impractical or hazardous in real-world scenarios. %It allows for simulation and data collection, 
This allows for in-depth testing of adversarial attack strategies and defensive measures without the  costs and risks of real-world testing.

%CARLA's detailed urban layouts, diverse vehicle models, and comprehensive sensor suites offer a versatile setting for experiments that would be impractical or hazardous in real-world scenarios. It offers a highly customized and controlled environment to perform simulation and data collection, % of our adversarial attacks. 
%enabling thorough analysis and testing of adversarial attack strategies and defensive measures without the %associated  costs and risks of real-world testing.

%Enhancing CARLA with the Robotic Operating System (ROS) \cite{ROS} forms our experiments' robust and modular framework. 

Integrating the Robotic Operating System (ROS) \cite{ROS} with CARLA creates a robust and modular framework for studying the effects of adversarial attacks. The CARLA-ROS-Bridge architecture %facilitates seamless communication between active nodes, 
ensures synchronized data exchange among all sensors and processing units via standardized messaging formats in communication protocols. This enables real-time operational testing of AV navigation systems' responses to adversarial attacks and the effectiveness of their countermeasures. The integration is essential for simulating complex attack patterns and thoroughly evaluating AV systems in real-world scenarios. % This integration is crucial for simulating complex attack patterns and real-time operational testing of AV navigation systems, enabling a comprehensive evaluation of system responses and countermeasures.
By leveraging %the capabilities of 
CARLA and ROS, our research aims to identify critical vulnerabilities in camera perception and devise robust strategies to bolster the cybersecurity of AVs. The insights gained from this research are imperative for ensuring AVs' safe and secure operation against evolving cyber threats. As we move towards a future where AVs are ubiquitous on our roads, our findings will contribute to developing more resilient AV systems, enhancing public trust and safety.

\begin{figure}[!t]
    \centering
    \includegraphics[width=0.85\linewidth]{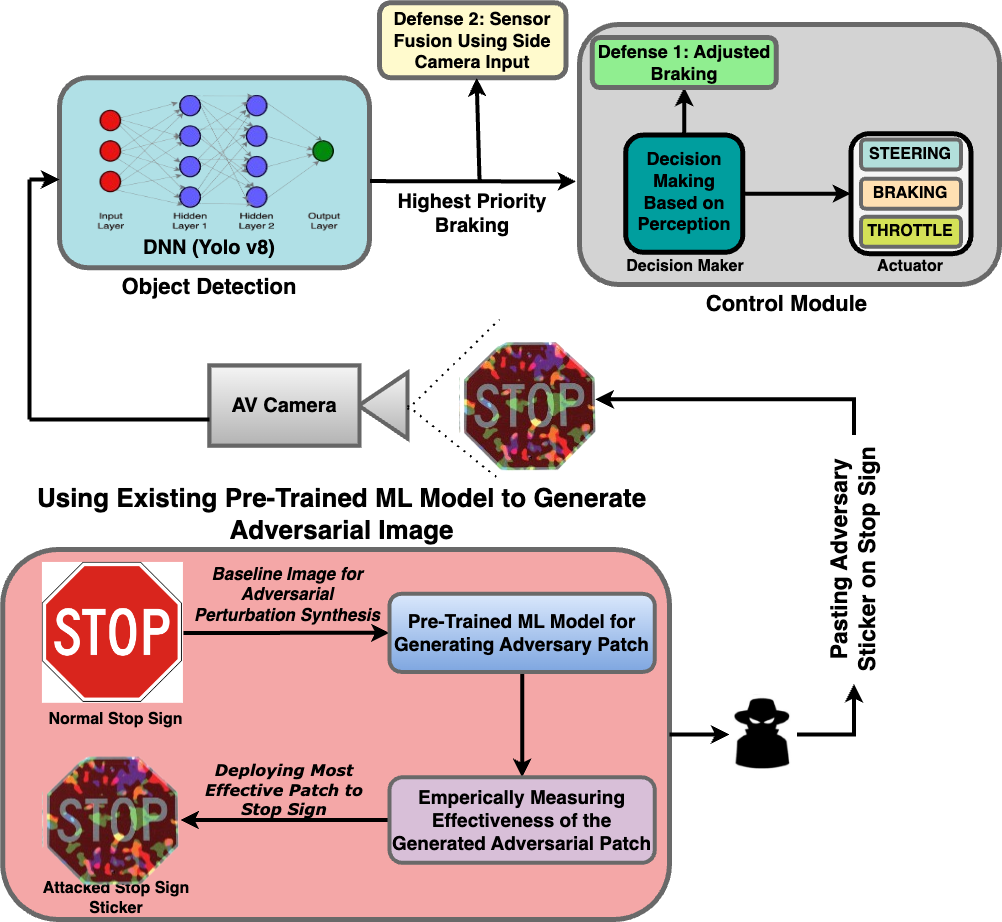}
    \caption{\small The architecture of the proposed adversarial attack with defense methods. %Architecture showing how an adversarial attack on stop sign detection in autonomous vehicles is performed and defended against. An adversarial patch is created using a pre-trained ML model and placed on a stop sign to mislead the object detection model (YOLO v8). Two defenses are used separately, Defense 1 (Adjusted Braking) and Defense 2 (Sensor Fusion with Side Camera).\newline \textbf{Note:} \textbf{Defense 1} and \textbf{Defense 2} are independent mechanisms and do not work simultaneously.
}
    \label{fig:ThreatModel}
    \vspace{-6mm}
\end{figure} 

\vspace{-2mm}
\section{Threat Model  and Overview of the Proposed Attack and Defense Approach}
\vspace{-1mm}
 Fig. \ref{fig:ThreatModel} shows the architecture of the proposed adversarial attack along with defense methods. In our threat model, an attacker aims to deceive the traffic sign detection model on AVs to disrupt their actuators, potentially causing safety issues. Specifically, the attacker generates an adversarial patch via existing adversarial image generation techniques \cite{chen2019shapeshifter, eykholt2018robust, lu2017adversarial, yang2020targeted}, applying it to a traffic sign to mislead the object detection model used by the AV camera, thereby affecting the AV’s actuating systems. To defend against such adversarial attacks, we have implemented two defense strategies: Defense 1 (adjusted braking), which adaptively adjusts braking system based on the AV's speed and distance to a detected stop sign (which is shown on the top right hand side of Fig. \ref{fig:ThreatModel}). Defense-2 which uses sensor fusion from a side camera. % applies a full brake once a stop sign is detected. %The threat arises when an attacker manipulates the appearance of the stop sign in a way that deceives the object detection system. %the AV operates in the CARLA simulator. The CARLA control module makes decisions based on camera perception of traffic signs and manages the actuators, including steering, braking, and throttle. The AV's object detection is performed using the YOLOv8x model \cite{yolo}, which allows the identification and response to traffic signs, such as stop signs. An attacker applies a camouflaged attack by manipulating the appearance of the stop sign to deceive the AV's object detection and disrupt its actuators. To defend against adversarial attacks, we have implemented two strategies. First, an adaptive braking system adjusts based on the AV's speed and distance to a detected stop sign. Second, a secondary camera applies a full brake once a stop sign is detected. %The threat arises when an attacker manipulates the appearance of the stop sign in a way that deceives the object detection system. 
Overall, this model focuses on these attack vectors and evaluates the effectiveness of the defensive measures in maintaining the AV's safety. 
\vspace{-2mm}
\section{%Sampling and Analysis of 
Methodology of Adversarial Attacks}% Initialization}
% In this section, we describe the methodology of adversarial attacks setup.
\vspace{-1mm}
\subsection{Image Modification Overview}
\vspace{-1mm}
\begin{figure}[!t]
    \centering
    \includegraphics[width=0.85\linewidth]{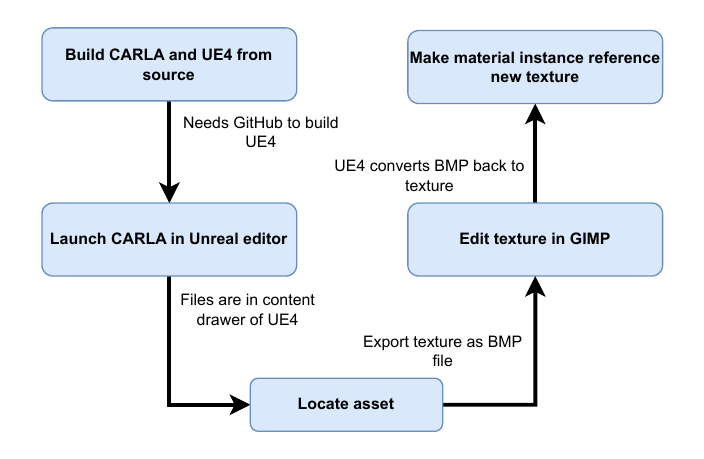}
    \vspace{-2mm}
    \caption{\small Methodology flow of image modification process.}
    \label{fig:flowchart}
     \vspace{-6mm}
\end{figure}

% General overview of how imagery textures were modified, types, and why.
% In order to simulate 
To mimic adversarial attacks in CARLA, %\cite{carla} 
the images of a stop sign asset are altered as described in Fig. \ref{fig:flowchart}. First, CARLA and Unreal Engine (UE) 4 \cite{unrealengine} have to be built from source in order to have direct access to the assets that need to be edited. %In order to access Unreal Engine 4 source you must have an active GitHub account. Once CARLA is opened in the Unreal Editor all assets used in the maps can be viewed and edited as desired. In order to change the image of the asset first 
The material instance of the object must be located in the content drawer of the Unreal Editor. Next, the referenced texture must be located and exported as a bitmap file. After exporting the texture, it is opened in GIMP (GNU Image Manipulation Program) \cite{gimp} and edited to the desired adversarial attack. The unique patterns %that were uniquely 
generated were inspired by research papers \cite{chen2019shapeshifter,eykholt2018robust,lu2017adversarial,yang2020targeted}. %Upon generating the adversarial attacks 
Once the adversarial images are created, the material instance of the stop sign must be updated to reference the adversarial image instead. %Once the asset is changed, 
A separate CARLA build is then made for each adversarial stop sign, facilitating easy and consistent testing for each attack. %The method used to alter the texture of stop signs T
%This method of altering the texture can also be applied to assets, such as cars and pedestrians. %Lastly, rather than directly change the texture of an asset, it makes more sense to add a sticker to the asset blueprint in the content drawer which can then be given any texture necessary to create an adversarial attack.
Additionally, instead of directly changing the texture of an asset, it is more efficient to add a sticker to the asset blueprint in the content drawer, allowing for any necessary texture to be applied to create an adversarial attack.

\begin{figure}[!t]
\center
\begin{subfigure}[b]{.2\textwidth} %48

	\includegraphics[width=\textwidth]{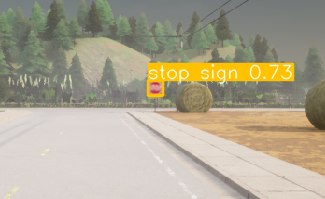}
	 \caption{\small  }
\label{fig:Non-attack}
 \end{subfigure}
%  \hfill
\begin{subfigure}[b]{.2\textwidth}

	\includegraphics[width=\textwidth]{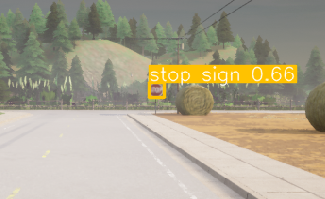}
	\caption{}
  \label{fig:Chen1}
 \end{subfigure}
 %  \hfill
  \vspace{2mm}
\begin{subfigure}[b]{.2\textwidth}

	\includegraphics[width=\textwidth]{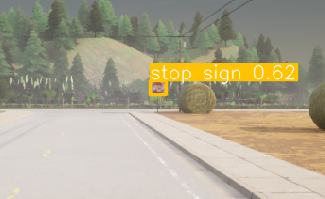}
	\caption{}
  \label{fig:Eykholt3}
 \end{subfigure}
  \vspace{-1mm}
 \begin{subfigure}[b]{.2\textwidth} %15

	\includegraphics[width=\textwidth]{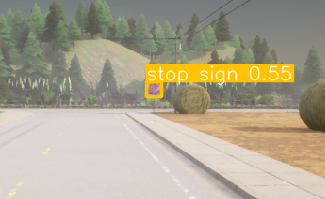}
	 \caption{\small  }
\label{fig:Lu2}
 \end{subfigure}
 \begin{subfigure}[b]{.21\textwidth}

	\includegraphics[width=\textwidth]{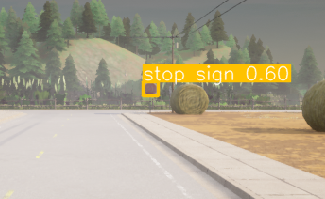}
	\caption{}
  \label{fig:Lu3}
 \end{subfigure}
 \begin{subfigure}[b]{.21\textwidth}

	\includegraphics[width=\textwidth]{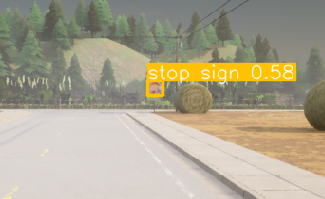}
	\caption{}
  \label{fig:Yang}
 \end{subfigure}
 %  \hfill
%  \hfill
  \vspace{-2mm}
 \caption{\small Stop sign detection performance with: (a) Non-attacked image, (b) Chen-modified image, (c) Eykholt-modified image, (d) Lu version-2 modified image, (e) Lu version-3 modified image, and (f) Yang-modified image.  }
\label{fig:yolo1}
  \vspace{-6mm}
\end{figure}

\vspace{-2mm}
\subsection{Attack Vectors and Effects}
\vspace{-1mm}
Different adversarial attacks are applied %to the AV camera 
to modify the texture of a stop sign image, and the impact of these attacks on object detection performance is analyzed. Initially, the object detection performance is tested on a non-attacked stop sign image, achieving a detection rate of $75\%$, as shown in Fig. \ref{fig:Non-attack}. Then, the stop sign image is attacked using Chen, Eykholt, Lu version-2, Lu version-3, and Yang adversarial attacks \cite{chen2019shapeshifter,eykholt2018robust,lu2017adversarial,yang2020targeted}. As illustrated in Fig. \ref{fig:Chen1}-\ref{fig:Yang}, all of these attacks result in a noticeable drop in object detection performance compared to the non-attacked image,  with Lu version-2 exhibiting the lowest detection score among all tested attacks. 

Following this, the impact of the attacks is tested on the AV's auto-braking system in CARLA. The AV was positioned at a distance that allowed it to reach the desired speed while keeping the stop sign outside the camera's object detection range. Once the AV reached the desired speed and the stop sign entered the detection range, the AV applied auto-braking to stop at the stop sign. It is observed that, without any attack on the stop sign image, the AV successfully performed a complete stop at the desired location. However, with the adversarial attacks on the stop sign image, the object detection performance degraded, causing a delayed auto-braking response. As a result, the AV failed to stop in time and passed the stop sign, leading to a potential safety issue.

%First, we determined the most effective adversarial attack on the stop signs by selecting the version that had the greatest impact on object detection. As shown in Figures \ref{fig:yolo1} and \ref{fig:yolo2}, there is a noticeable drop in detection performance between the normal stop sign on Figure 2 and the adversarially attacked stop signs on Figure 3. Specifically, the Lu version 2 attack in Figure \ref{fig:yolo2} exhibits the lowest object detection score among all tested variations. We then set up the testing environment by positioning the vehicle at a distance that allowed it sufficient time to reach the desired speed while keeping the stop sign outside the camera's object detection range. We also tested various braking parameters on a normal stop sign to ensure the vehicle would come to a complete stop at the desired position. Once the optimal parameters were achieved we applied them onto the simulation with the adversarial stop sign so that we could evaluate their impact on the object detection and how this affected the vehicle’s braking performance.
\vspace{-3mm}
\subsection{Mitigating Attack Vectors}
\vspace{-1mm}
\begin{comment}
    To counteract the adversarial attacks in the CARLA simulator, we implemented two defense strategies. The first defense involved calculating the distance from the stop sign using object detection as well as using the vehicle’s velocity to ensure the vehicle would stop at the desired location. The distance formula was calculated by taking in the bounding box height from the object detection, the height of the stop sign, camera’s field of view and the image height to calculate the focal length, which was then used to calculate the distance. With the estimated distance, we then calculated a braking force based on the vehicle’s speed and the distance from the stop sign. The second defense strategy involved implementing a side camera that applies a full brake once the stop sign is detected. This option provides an additional layer of safety so that in the event that the frontal camera fails the side camera can take over and stop the vehicle.
\end{comment}

%{\color{red}Equations in the following need to be explained or provide citations.}
To address the vulnerabilities exposed by adversarial stop signs, we propose two defense methods: 1) Defense-1 (adjusted braking) and 2) Defense-2 (sensor fusion using a side camera).  
\subsubsection{Adjusted Braking}
This strategy involves adjusting the AV’s braking based on the distance from the stop sign and its velocity. It calculates the distance using object detection and takes the AV’s velocity into account to ensure that the AV stops at the desired location. To do so,  %our first defense involved adjusting the distance parameters so that the object detection is able to detect the stop sign in both scenarios before the braking is applied. Initially, t
we calculate the focal length in pixels \( f \) using the formula \cite{szeliski2010}: 
\vspace{-1.5mm}
\begin{equation}
f = \frac{H_{image}}{2 \cdot \tan\left(\frac{\text{FOV}}{2}\right)}
\end{equation}
where $H_{image}$ is the height of the camera sensor in pixels and $\text{FOV}$ is the field of view of the camera in degrees. Using the focal length, the distance $d$ to the stop sign is calculated as follows \cite{szeliski2010}:
\vspace{-0.8mm}
\begin{equation}
d = \frac{H_{real} \cdot f}{h_{bbox}}
\vspace{-0.5mm}
\end{equation}
where $H_{real}$ is the real height of the stop sign and $h_{bbox}$ is the height of the bounding box in pixels. After that, we adjust the braking system to account for both the current distance from the stop sign and the AV's velocity, enabling a more adaptive response and ensuring realistic testing conditions. Two additional parameters can be calculated in the following: the  deceleration \( a \) required  to stop the AV, and the brake control message \( b \) that indicates the required braking force. As described in \cite{serway2018}, the  deceleration \( a \) is given by:%using the formula:
\vspace{-1.8mm}
\begin{equation}
a = \frac{v^2}{2 \cdot d_{stop}}
\vspace{-0.5mm}
\end{equation}

where \( v \) is the AV speed and \( d_{stop} \) is the distance to the stop sign. The brake control message \( b \) can be calculated as follows:%using the formula given in \cite{szeliski2010}:
\vspace{-0.5mm}
\begin{equation}
b = \min\left(\frac{a}{a_{max}} \cdot M , ~1\right)
\vspace{-0.5mm}
\end{equation}
where \( a_{max} \) is the reference maximum deceleration that a typical vehicle can safely apply under ideal conditions, usually set to $9.81 m/s^2$, $M$ is the brake multiplier, which adjusts the applied braking force and is typically set between $0.6$ and $1$. Note that $b$ ranges form $0$ to $1$, where $0$ indicates no braking and $1$ indicates full braking.

\subsubsection{Sensor Fusing Using a Side Camera}
This strategy leverages  a side camera on the AV to apply full braking once the stop sign is detected, adding an extra layer of safety. This approach ensures that if the frontal camera is compromised, the side camera can take over and stop the AV.

%Additionally, another defense involved implementing a side camera that applies a full brake once a stop sign is detected. This method was tested using the original parameters for the attack mentioned in the previous part. By having a secondary camera take over, this approach ensures an additional layer of safety if the primary detection system is compromised.
%Both methods successfully defended against this type of adversarial attack. %This ensures that if the primary detection system were to be attacked there would still be another layer of safety by having a secondary camera take over. Both methods mentioned were successful in defending against this type of adversarial attack.

\vspace{-2mm}
\section{Assessment Results}
\vspace{-1mm}
In this section, we evaluate the effectiveness of the proposed adversarial attack and the corresponding mitigation strategies. Notably, we focus on the Lu adversarial attack, as it results in the most significant decline in object detection performance. 
%Once the confidence results for each adversarial stop sign were analyzed, we focused on testing the one based on the Lu method due to there being a larger confidence drop compared to the other methods.

\vspace{-1.5mm}
\subsection{Adversarial Stop Sign Attacks}
\begin{figure}[!t]
\center
\begin{subfigure}[b]{.19\textwidth}

	\includegraphics[width=\textwidth]{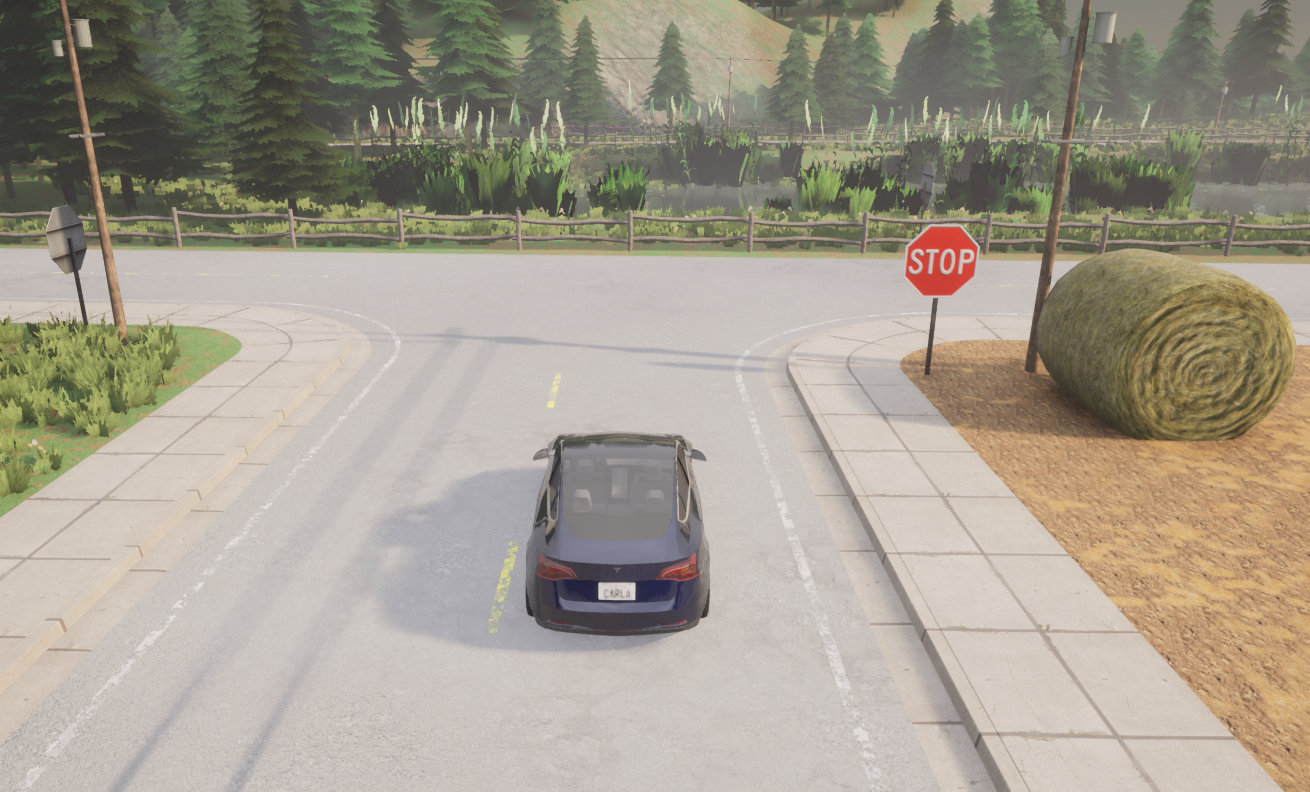}
	 \caption{\small  }
\label{fig:normal-stop}
 \end{subfigure}
%  \hfill
\begin{subfigure}[b]{.19\textwidth}

	\includegraphics[width=\textwidth]{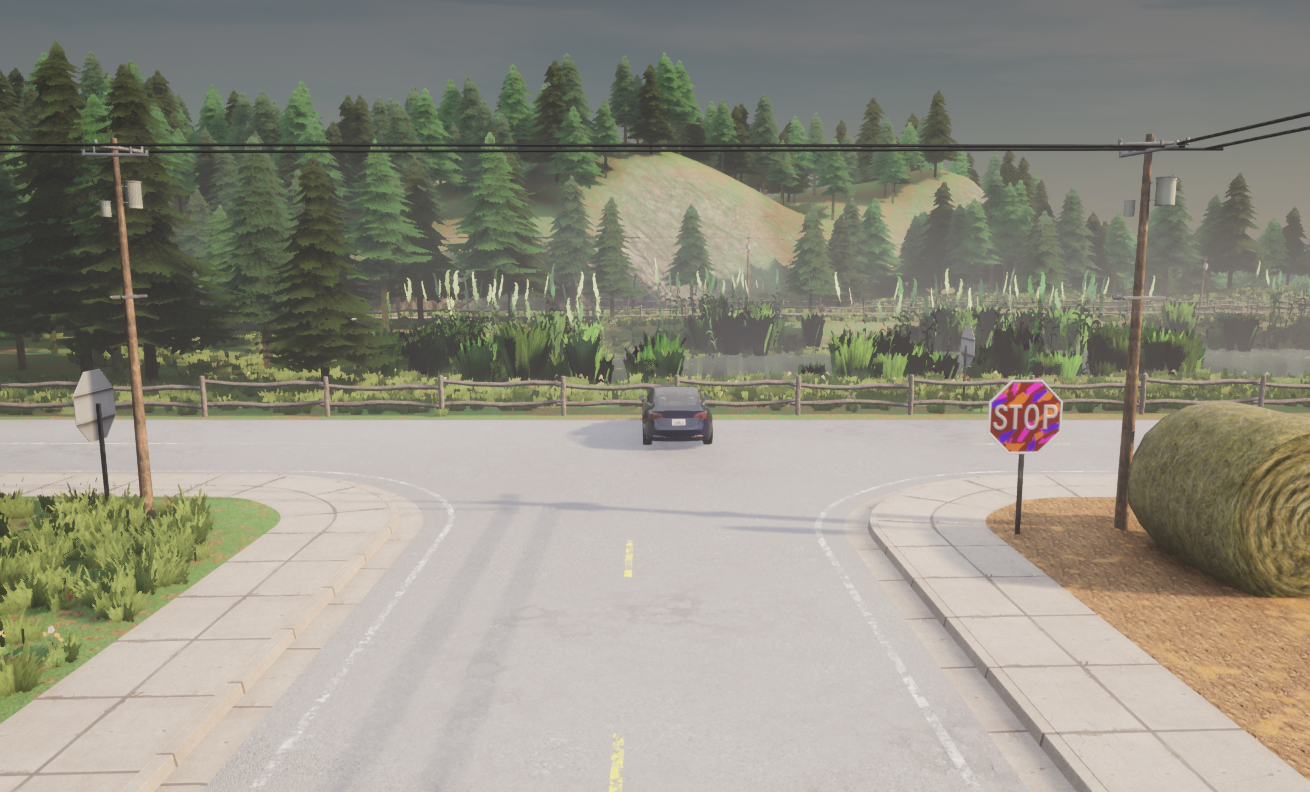}
	\caption{}
  \label{fig:attacked-stop}
 \end{subfigure}
 %  \hfill
  \vspace{-2mm}
 \caption{\small AV's auto-braking response with: (a) non-attacked stop sign, (b) attacked stop sign, both positioned in standard location in one of the provided maps of CARLA (Town 07). Note that vehicle didn't stop at the stop sign.}
 \label{fig:attack}
  \vspace{-3.5mm}
\end{figure}

\vspace{-1mm}
The testing procedure involves evaluating an AV's auto-braking to both non-attacked and attacked stop sign image in the CARLA simulator. %placing the CARLA AV in different towns, with normal and adversarial stop signs positioned at varying distances from the road, categorized as standard, far, and near. 
During the test, the AV  accelerates to a specified speed ($85$ km/h), maintaining that speed until a stop sign is detected. Once the stop sign is detected with a confidence above a certain threshold, constant braking is applied.  Fig. \ref{fig:normal-stop} shows that the AV successfully stopped at the intended position when the non-attacked stop sign was in place.  However, when an attack was applied to the stop sign, the AV passed the stop sign without stopping. Fig. \ref{fig:attacked-stop} illustrates the final location of the AV when faced with the adversarially attacked stop sign. This finding highlights the significant impact of adversarial attacks on AV behavior, underscoring the necessity for effective countermeasures.

\begin{figure}[!t]
\center
\begin{subfigure}[b]{.17\textwidth}

	\includegraphics[width=\textwidth]{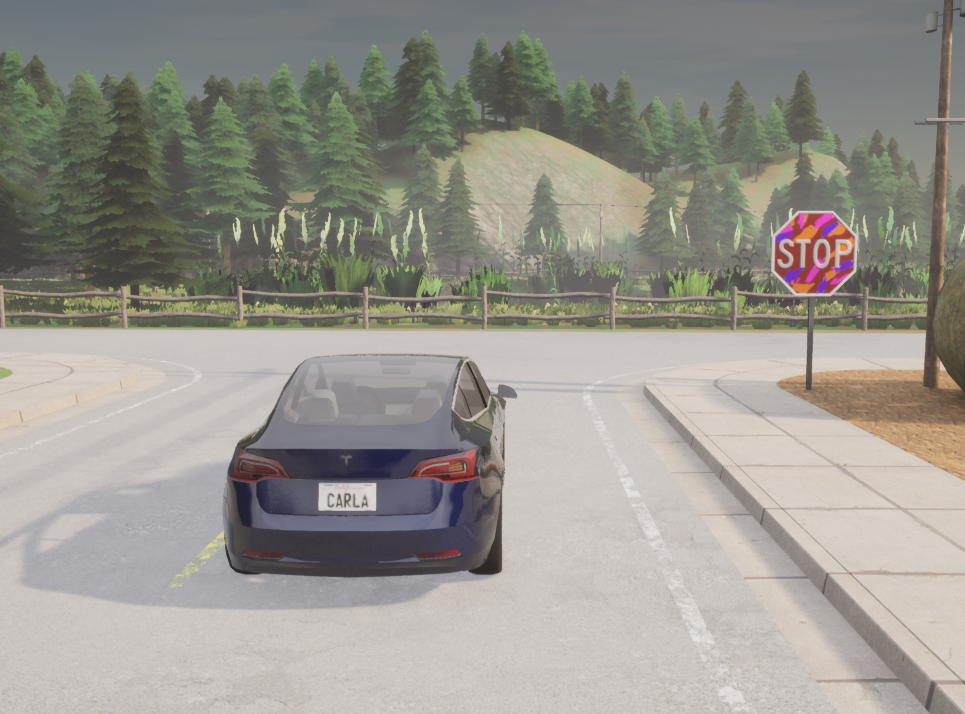}
	 \caption{\small  }
\label{fig:def-1}
 \end{subfigure}
%  \hfill
\begin{subfigure}[b]{.17\textwidth}

	\includegraphics[width=\textwidth]{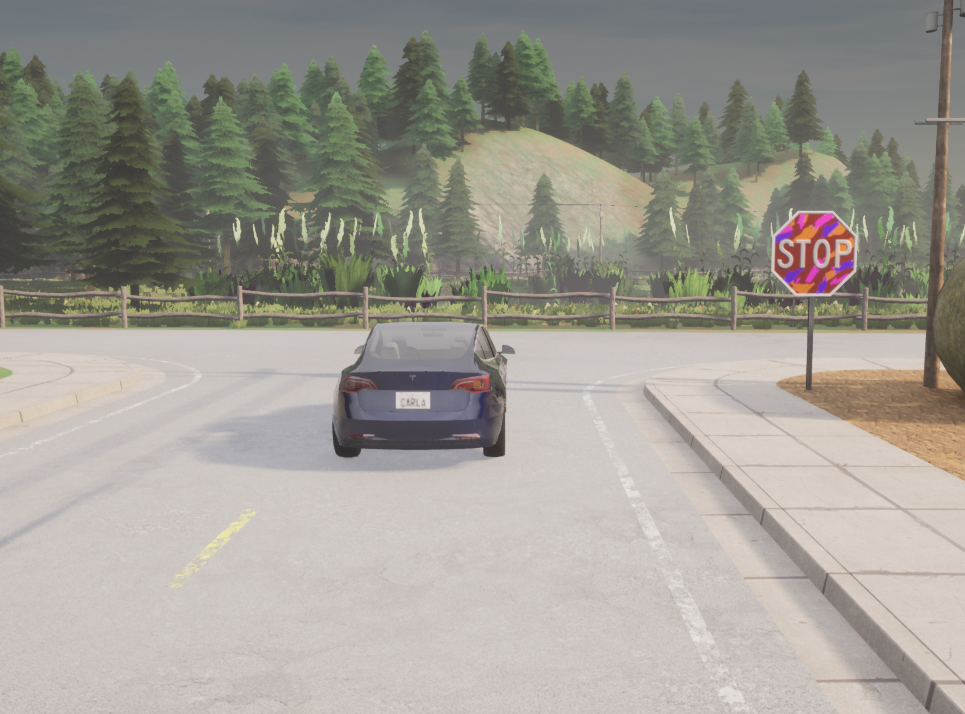}
	\caption{}
  \label{fig:def-2}
 \end{subfigure}
 %  \hfill
  \vspace{-1mm}
 \caption{\small AV's auto-braking response to an attacked stop sign after applying: (a) Defense-1 strategy, (b) Defense-2 strategy.  }
 \label{fig:Defense}
  \vspace{-2mm}
\end{figure}

\setlength\extrarowheight{4pt}
\begin{table}[!t]
\caption{\small AV's stopping positions under different test scenarios.} %\\ The location of the stop sign is (-9.3, -50.5)}
\begin{center}
\vspace{-3mm}
\begin{tabular}{|c|c|}
\hline
\textbf{Scenario} & \textbf{AV stopping coordinates (X,Y) in meters} \\
\hline
Normal Stop Sign &  (-5.4, -46.2)\\
\hline
Attacked Stop Sign & (-5.4, -60.8) \\
\hline
Adjusted Braking & (-5.4, -43.5) \\
\hline
Side Camera & (-5.4, -47.5) \\
\hline
CARLA Autopilot & (-5.9, -44.5) \\
\hline
\end{tabular}
\label{tab3}
\end{center}
\vspace{-6mm}
\end{table}

\vspace{-3mm}
\subsection{Adversarial Stop Sign Mitigation Strategies}
\vspace{-1.5mm}
To validate the effectiveness of the implemented defense strategies, we conducted a series of tests to compare the AV's stopping locations %under different conditions: 
during an adversarial attack and after applying our defense strategies. %, and using the CARLA autopilot mode. %The results highlight a significant improvement in stopping accuracy when our defense mechanisms are employed.

%\subsubsection{%Braking Adjusted Based on Stop Sign Distance and Velocity 
%Adjusted Braking} This strategy involves adjusting the AV's braking based on the distance from the stop sign and its velocity. It calculates the distance using object detection and takes the AV’s velocity into account to ensure that the AV stops at the desired location. 
As given in Table \ref{tab3}, during the attack, the AV’s final position was approximately at $(-5.4,-60.8)$, which is $10$ m beyond the stop sign located at $(-9.3, -50.5)$. With the adjusted braking defense strategy, %which takes into account the AV's current speed and distance from the stop sign, 
we successfully reduced the detection distance required for effective braking, ensuring  the AV stop a few meters before the stop sign. The final stopping position is illustrated in Fig. \ref{fig:def-1}. Using the side camera, the final stopping position %of the AV 
 was $(-5.4,-47.5)$, a few meters before the stop sign, as given in Table \ref{tab3} and  depicted in Fig. \ref{fig:def-2}. These results confirm that both defense methods are effective in mitigating adversarial stop sign attacks, ensuring that the AV stops at a safe and accurate location.

%\subsubsection{ Sensor Fusing Using a Side Camera}
%This strategy utilizes a side camera on the AV to apply full braking once the stop sign is detected. This method provides an additional layer of safety, ensuring that if the frontal camera fails, the side camera can take over and stop the AV. 

\begin{figure}[!t]
\center
\begin{subfigure}[b]{.21\textwidth}

	\includegraphics[width=\textwidth]{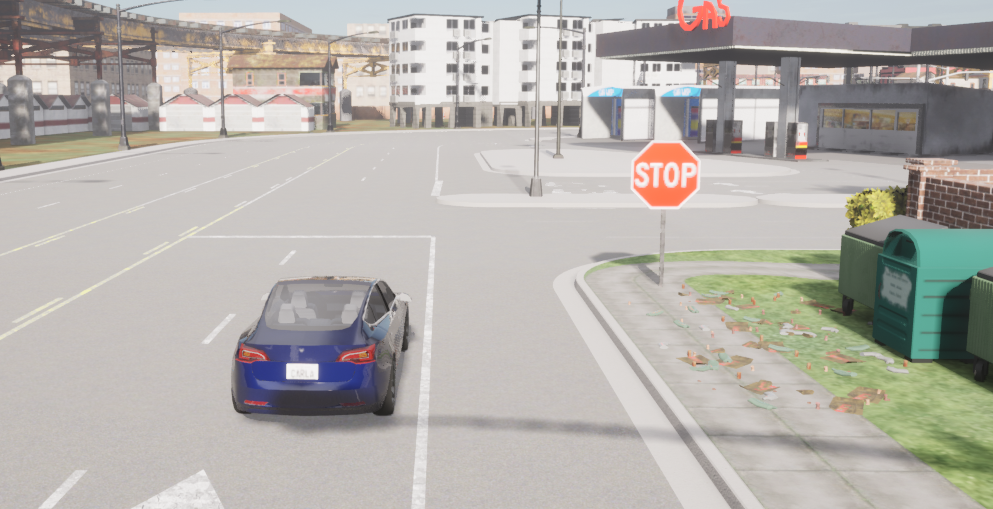}
	 \caption{\small  }
\label{fig:near-stop}
 \end{subfigure}
%  \hfill
\begin{subfigure}[b]{.21\textwidth}

	\includegraphics[width=\textwidth]{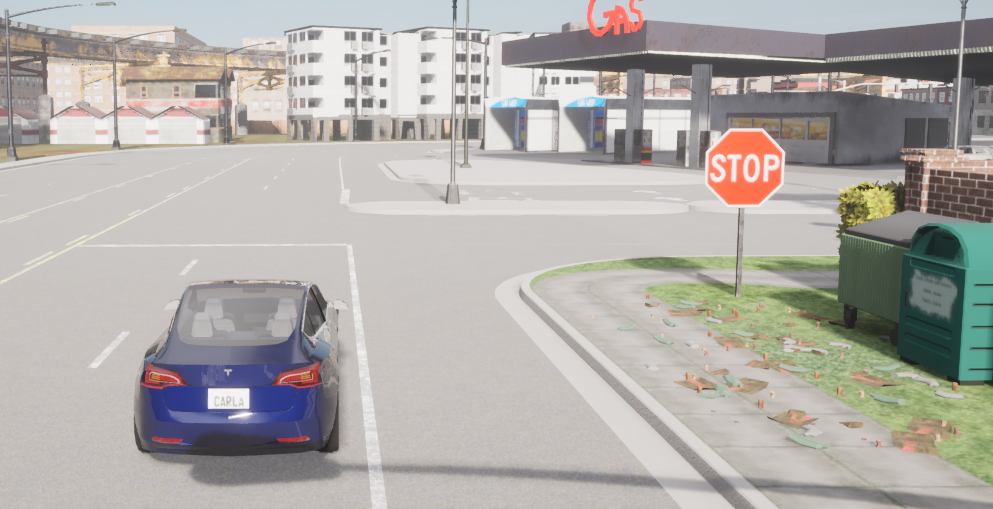}
	\caption{}
  \label{fig:far-stop}
 \end{subfigure}
 %  \hfill
  \vspace{-2.5mm}
 \caption{\small AV's auto-braking response to non-attacked stop sign positioned: (a)  near the road, (b) far from the road, in one of the provided maps of CARLA (Town 03). Notice the location of the stop sign with respect to the road.   }
 \label{fig:near-far}
  \vspace{-3mm}
\end{figure}

Furthermore, the effectiveness of these methods has been tested by placing the CARLA AV in different towns (maps provided by CARLA simulator), with normal and adversarial stop signs positioned at varying distances from the road, categorized as \textbf{standard}, \textbf{far}, and \textbf{near}. Various parameters, such as the time to complete stop and the distance to stop sign when brake is applied, are measured. %we tested the impact of changing  the stop sign's location across different towns, considering  the non-attacked scenario. 
Fig. \ref{fig:near-far} shows that the AV comes to a complete stop when encountering a non-attacked stop sign, regardless of its location—standard, near, or far from the road—in different towns after applying our  combined defense strategies. Additionally, Table \ref{tab1} provides the time to complete stop and the distance to the non-attacked  stop sign. The values indicates that, regardless of the stop sign's placement, the AV can detect it and come to a complete stop. A similar test with attacked stop signs revealed that our defense strategies reliably detect the stop sign and applies the brakes, countering the proposed attack, as given in Table \ref{tab2}. These findings %images and coordinates 
demonstrate the robustness of our defense strategies in maintaining the AV's stopping performance, even under adversarial conditions, with consistent results across varied environments and stop sign placements.

%the adversarial attack is the only factor significantly impacting the AV's braking behavior, irrespective of the stop sign’s location. These findings highlight the significant impact of adversarial attacks on AV behavior, underscoring the necessity for effective countermeasures. 

%attacked stop signs in different towns and at varying distances from the road. As given in Table \ref{tab2},  our approach reliably detects the stop sign and applies the brakes, countering the proposed attack.  The findings %images and coordinates 
%demonstrate the robustness of our defense strategies in maintaining the AV's stopping performance, even under adversarial conditions.

\setlength\extrarowheight{4pt}
\begin{table}[!t]
\caption{\small AV's collected data in response to a non-attacked stop sign located in different CARLA towns after applying both proposed defense methods.}
\vspace{-3mm}
\begin{center}
\begin{tabular}{|c|c|c|}
\hline
 \multirow{2}{*}{\textbf{Map / stop }} &  \multirow{2}{*}{\textbf{Time to complete}} &  \multirow{2}{*}{\textbf{Distance to stop  }} \\
\textbf{sign position}& \textbf{stop (s)} & \textbf{ sign %when brake is applied 
(m)} \\
\hline
Town-07 (Standard) & 2.70 & 39.12 \\
\hline
Town-03 (Far) & 2.60 & 35.95 \\
\hline
Town-03 (Near) & 2.71 & 33.26 \\
\hline
Town-10 (Far) & 2.75 & 39.12 \\
\hline
Town-10 (Near) & 2.75 & 39.12 \\
\hline
\end{tabular}
\label{tab1}
\end{center}
\vspace{-3.5mm}
\end{table}

\setlength\extrarowheight{4pt}
\begin{table}[!t]
\caption{\small AV's collected data in response to an attacked stop sign located in different CARLA towns after applying both proposed defense methods.}
\vspace{-3mm}
\begin{center}
\begin{tabular}{|c|c|c|}
\hline
 \multirow{2}{*}{\textbf{Map / stop }} &  \multirow{2}{*}{\textbf{Time to complete}} &  \multirow{2}{*}{\textbf{Distance to stop  }} \\
\textbf{sign position}& \textbf{stop (s)} & \textbf{ sign %when brake is applied 
(m)} \\
\hline
Town07 (Standard) & 2.70 & 38.01 \\
\hline
Town03 (Far) & 2.55 & 25.58 \\
\hline
Town03 (Near) & 2.75 & 35.01 \\
\hline
Town10 (Far) & 2.70 & 35.01 \\
\hline
Town10 (Near) & 2.70 & 35.01 \\
\hline
\end{tabular}
\label{tab2}
\end{center}
\vspace{-6mm}
\end{table}

%attacked stop signs in different towns and at varying distances from the road. As given in Table \ref{tab2},  our approach reliably detects the stop sign and applies the brakes, countering the proposed attack.  The findings %images and coordinates 
%demonstrate the robustness of our defense strategies in maintaining the AV's stopping performance, even under adversarial conditions.

\begin{comment}
\begin{figure} [H]
    \centering
    \includegraphics[width=0.4\linewidth]{Defense1.png}
    \includegraphics[width=0.4\linewidth]{Defense2.png}
    \caption{Left Image includes the stopping position for Defense 1 \\ Right Image includes the stopping position for Defense 2}
    \label{fig:Defense}
\end{figure}    
\end{comment}

\vspace{-2mm}
\section{Conclusions}
\vspace{-2mm}
We introduced a sophisticated framework for attacking AV camera sensors and deceiving AI models, demonstrating its effectiveness on the CARLA simulator. The attack focuses on targeting traffic sign recognition (TSR) to manipulate AV actuators. %(i.e., steering, braking, and throttle). 
In particular, the attack modifies the texture of a stop sign to fool the AV's object detection model, delaying the braking system and potentially causing safety issues. Following this, we presented and tested effective mitigation strategies that can be applied as countermeasures against adversarial attacks on TSR, ensuring AV safety. The comprehensive analysis of the proposed attack and defense methods offers valuable insights into securing and ensuring the safe operation of AVs, especially in a zero-trust environment.   %applicable to other end-to-end trained autonomous cyber-physical systems. %The proposed attack and defense methods are also applicable to other end-to-end trained autonomous cyber-physical systems.    
\bibliographystyle{IEEEtran}
\bibliography{testref}

% Generated by IEEEtran.bst, version: 1.12 (2007/01/11)
\begin{thebibliography}{10}
\providecommand{\url}[1]{#1}
\csname url@samestyle\endcsname
\providecommand{\newblock}{\relax}
\providecommand{\bibinfo}[2]{#2}
\providecommand{\BIBentrySTDinterwordspacing}{\spaceskip=0pt\relax}
\providecommand{\BIBentryALTinterwordstretchfactor}{4}
\providecommand{\BIBentryALTinterwordspacing}{\spaceskip=\fontdimen2\font plus
\BIBentryALTinterwordstretchfactor\fontdimen3\font minus \fontdimen4\font\relax}
\providecommand{\BIBforeignlanguage}[2]{{%
\expandafter\ifx\csname l@#1\endcsname\relax
\typeout{** WARNING: IEEEtran.bst: No hyphenation pattern has been}%
\typeout{** loaded for the language `#1'. Using the pattern for}%
\typeout{** the default language instead.}%
\else
\language=\csname l@#1\endcsname
\fi
#2}}
\providecommand{\BIBdecl}{\relax}
\BIBdecl

\bibitem{kloukiniotis2022countering}
A.~Kloukiniotis, A.~Papandreou, A.~Lalos, P.~Kapsalas, D.-V. Nguyen, and K.~Moustakas, ``Countering adversarial attacks on autonomous vehicles using denoising techniques: A review,'' \emph{IEEE Open Journal of Intelligent Transportation Systems}, vol.~3, pp. 61--80, 2022.

\bibitem{mahima2024toward}
K.~Y. Mahima, A.~G. Perera, S.~Anavatti, and M.~Garratt, ``Toward robust 3{D} perception for autonomous vehicles: A review of adversarial attacks and countermeasures,'' \emph{IEEE Transactions on Intelligent Transportation Systems}, 2024.

\bibitem{parekh2022review}
D.~Parekh, N.~Poddar, A.~Rajpurkar, M.~Chahal, N.~Kumar, G.~P. Joshi, and W.~Cho, ``A review on autonomous vehicles: Progress, methods and challenges,'' \emph{Electronics}, vol.~11, no.~14, p. 2162, 2022.

\bibitem{sarwatt2024adapting}
D.~S. Sarwatt, F.~Kulwa, J.~Ding, and H.~Ning, ``Adapting image classification adversarial detection methods for traffic sign classification in autonomous vehicles: A comparative study,'' \emph{IEEE Transactions on Intelligent Transportation Systems}, 2024.

\bibitem{triki2023real}
N.~Triki, M.~Karray, and M.~Ksantini, ``A real-time traffic sign recognition method using a new attention-based deep convolutional neural network for smart vehicles,'' \emph{Applied Sciences}, vol.~13, no.~8, p. 4793, 2023.

\bibitem{pavlitska2023adversarial}
S.~Pavlitska, N.~Lambing, and J.~M. Z{\"o}llner, ``Adversarial attacks on traffic sign recognition: A survey,'' in \emph{Proc. of the IEEE 3rd International Conference on Electrical, Computer, Communications and Mechatronics Engineering (ICECCME)}, 2023.

\bibitem{soylu2024performance}
E.~Soylu and T.~Soylu, ``A performance comparison of {YOLO}v8 models for traffic sign detection in the {R}obotaxi-full scale autonomous vehicle competition,'' \emph{Multimedia Tools and Applications}, vol.~83, no.~8, pp. 25\,005--25\,035, 2024.

\bibitem{jia2022fooling}
W.~Jia, Z.~Lu, H.~Zhang, Z.~Liu, J.~Wang, and G.~Qu, ``Fooling the eyes of autonomous vehicles: Robust physical adversarial examples against traffic sign recognition systems,'' \emph{arXiv preprint arXiv:2201.06192}, 2022.

\bibitem{zhong2022shadows}
Y.~Zhong, X.~Liu, D.~Zhai, J.~Jiang, and X.~Ji, ``Shadows can be dangerous: Stealthy and effective physical-world adversarial attack by natural phenomenon,'' in \emph{Proc. of the IEEE/CVF Conference on Computer Vision and Pattern Recognition}, 2022.

\bibitem{sato2023wip}
T.~Sato, S.~H. Bhupathiraju, M.~Clifford, T.~Sugawara, Q.~A. Chen, and S.~Rampazzi, ``W{IP}: Infrared laser reflection attack against traffic sign recognition systems,'' in \emph{Porc. of ISOC Symposium on Vehicle Security and Privacy (VehicleSec)}, 2023.

\bibitem{carla}
C.~Team, ``Carla,'' \url{https://carla.org/}, 2023.

\bibitem{ROS}
``Robot operating system,'' \url{https://www.ros.org/}, 2019.

\bibitem{chen2019shapeshifter}
S.-T. Chen, C.~Cornelius, J.~Martin, and D.~H. Chau, ``Shapeshifter: Robust physical adversarial attack on faster {R}-{CNN} object detector,'' in \emph{Proc. of the European Conference on Machine Learning and Knowledge Discovery in Databases: European Conference (ECML PKDD)}.\hskip 1em plus 0.5em minus 0.4em\relax Springer, 2019.

\bibitem{eykholt2018robust}
K.~Eykholt, I.~Evtimov, E.~Fernandes, B.~Li, A.~Rahmati, C.~Xiao, A.~Prakash, T.~Kohno, and D.~Song, ``Robust physical-world attacks on deep learning visual classification,'' in \emph{Proc. of the IEEE conference on computer vision and pattern recognition}, 2018.

\bibitem{lu2017adversarial}
J.~Lu, H.~Sibai, and E.~Fabry, ``Adversarial examples that fool detectors,'' \emph{arXiv preprint arXiv:1712.02494}, 2017.

\bibitem{yang2020targeted}
X.~Yang, W.~Liu, S.~Zhang, W.~Liu, and D.~Tao, ``Targeted attention attack on deep learning models in road sign recognition,'' \emph{IEEE Internet of Things Journal}, vol.~8, no.~6, pp. 4980--4990, 2020.

\bibitem{unrealengine}
E.~Games, ``Unreal engine,'' \url{https://www.unrealengine.com/}, 2020.

\bibitem{gimp}
G.~Team, ``Gimp,'' \url{https://www.gimp.org/}, 2024.

\bibitem{szeliski2010}
R.~Szeliski, \emph{Computer Vision: Algorithms and Applications}.\hskip 1em plus 0.5em minus 0.4em\relax London: Springer, 2010.

\bibitem{serway2018}
R.~A. Serway and J.~W. Jewett, \emph{Physics for Scientists and Engineers}, 10th~ed.\hskip 1em plus 0.5em minus 0.4em\relax Cengage Learning, 2018.

\end{thebibliography}


\begin{thebibliography}{00}
\bibitem{b1} G. Eason, B. Noble, and I. N. Sneddon, ``On certain integrals of Lipschitz-Hankel type involving products of Bessel functions,'' Phil. Trans. Roy. Soc. London, vol. A247, pp. 529--551, April 1955.
\bibitem{b2} J. Clerk Maxwell, A Treatise on Electricity and Magnetism, 3rd ed., vol. 2. Oxford: Clarendon, 1892, pp.68--73.
\bibitem{b3} I. S. Jacobs and C. P. Bean, ``Fine particles, thin films and exchange anisotropy,'' in Magnetism, vol. III, G. T. Rado and H. Suhl, Eds. New York: Academic, 1963, pp. 271--350.
\bibitem{b4} K. Elissa, ``Title of paper if known,'' unpublished.
\bibitem{b5} R. Nicole, ``Title of paper with only first word capitalized,'' J. Name Stand. Abbrev., in press.
\bibitem{b6} Y. Yorozu, M. Hirano, K. Oka, and Y. Tagawa, ``Electron spectroscopy studies on magneto-optical media and plastic substrate interface,'' IEEE Transl. J. Magn. Japan, vol. 2, pp. 740--741, August 1987 [Digests 9th Annual Conf. Magnetics Japan, p. 301, 1982].
\bibitem{b7} M. Young, The Technical Writer's Handbook. Mill Valley, CA: University Science, 1989.
\end{thebibliography}

\end{document}